# SENVM: Server Environment Monitoring and Controlling System for a Small Data Center Using Wireless Sensor Network

Supasate Choochaisri, Vit Niennattrakul, Saran Jenjaturong,
Chalermek Intanagonwiwat, Chotirat Ann Ratanamahatana
Department of Computer Engineering, Faculty of Engineering,
Chulalongkorn University, Bangkok 10330, Thailand
Email: {g52sch, g49vnn, g49sjn, intanago, ann}@cp.eng.chula.ac.th

**Abstract**

*In recent years, efficient energy utilization becomes an essential requirement for data centers, especially in data centers of world-leading companies, where "Green Data Center" defines a new term for an environment-concerned data center. Solutions to change existing a data center to the green one may vary. In the big company, high-cost approaches including re-planning server rooms, changing air-conditioners, buying low-powered servers, and equipping sophisticating environmental control equipments are possible, but not for small to medium enterprises (SMEs) and academic sectors which have limited budget. In this paper, we propose a novel system, SENVM, used to monitor and control air temperature in a server room to be in appropriate condition, not too cold, where very unnecessary cooling leads to unnecessary extra electricity expenses, and also inefficient in energy utilization. With implementing on an emerging technology, Wireless Sensor Network (WSN), Green Data Center is feasible to every small data center with no wiring installation, easy deployment, and low maintenance fee. In addition, the prototype of the system has been tested, and the first phase of the project is deployed in a real-world data center.*

## 1. Introduction

In recent years, many world-leading organizations including governments and private sectors have been focusing on energy utilization, especially on their data centers, where each data center consumes energy nearly to a whole small village. "Green Data Center" [3] defines a new term represented an environmental-concerned data center which is well managed in energy consumption. Meanwhile several world gigantic companies, e.g., Google [21], and IBM [3], facilitate their own data centers to be green with high-cost solutions. In contrast, small to medium enterprises (SMEs) and academic sectors, which have limited budge, cannot afford any costly sophisticating monitoring and controlling equipment. Specifically, solutions for gigantic companies can be even to change to newer servers or air-conditioning systems, but clearly it is too costly for others, especially most of the data centers in Thailand.

Since Thailand is located in tropical region with average temperature varying from 23ºC to 30ºC in Central Province [19], all server rooms are required to be cooled by air-conditioning systems. Therefore, several organizations set their default temperature to be lower than 20ºC [1] for a reason to prevent servers over-heated from the highest load time. In other words, air-conditioner systems cannot detect 'real' temperature at a server; therefore, setting very low temperature may guarantee that every server is cold enough since air-conditioners. However, the American Society of Heating Refrigerating and Air-Conditioning Engineers (ASHRAE) states that the appropriate temperature in a server room should be up to 27ºC [17]. This means almost data centers over-consume energies which cause extra charges for electricity, nearly 3.4 million Baht a year for a small server room (containing around 20 250kW servers), as shown in Figure 1.

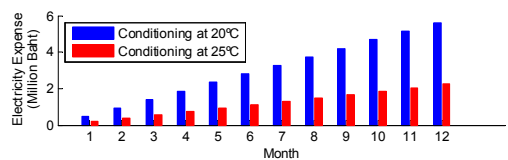

Figure 1. With temperature at 25ºC, electricity expenses are reduced approximately around 60% [8] or 3.4 million Baht for a small data center.

---
[1] By surveying many data centers in various organizations.

To make a solution feasible for a small server room resided in high-temperature climate, in this paper, we propose a system, called SENVM, for real-time temperature monitoring and controlling using Wireless Sensor Network (WSN) [1]. WSN is a promising technology that has several advantages such as low hardware and installation cost, ease of deployment, and very low maintenance fee. Our system dynamically adjusts air temperature to be in appropriate condition for servers. Instead of decreasing the same temperature of every single air conditioner, SENVM can detect a hot zone and decrease temperature in that specific area. As a result, not only the energy consumption of the cooling system will be significantly reduced, but the organizations will also remarkably cut down the electricity expense. With the WSN technology, organizations can effortlessly install the system with no additional wiring installation and maintenance cost. Additionally, our SENVM system is implemented to view the system logs and status via the Internet in real time.

The rest of paper is organized as follows. Section 2 provides related work and essential background on Wireless Sensor Network. We describe our SENVM system in Section 3, and demonstrate deployment of SENVM in Section 4. Finally, conclusion and future research direction are described in Section 5.

## 2. Related Work and Background

A wireless sensor network [1] consists of distributed tiny wireless sensing devices with some computational, storage and communication capability. These devices cooperatively form a network and communicate with each other to monitor and gather physical data from their surroundings, such as temperature, humidity, light intensity, pressure and vibration.

Each sensing device is typically small and easy to deploy because there is no need for communication wiring. Therefore, a WSN is well suited for varieties of applications. Previously, WSNs have been applied into several works to demonstrate how WSNs support and solve existing problems.

Mainwaring et al. [10] design an experimental WSN system deployed in the Great Duck Island for habitat monitoring. Their work monitors the nesting behavior of Leach's Storm Petrel and collects data into the central database. This is the very first experimental system to demonstrate how WSN can be applied to collect data from inaccessible fields.

Gui and Mohapatra [6] present their study of applying WSN as a target tracking application, while, He et al. [7] present a surveillance system using WSN. Both two works share the similar idea to detect and track moving objects. This kind of application can be used as a security application.

A structural monitoring application has been presented in Wisden [15]. The challenge of this work is similar to our system that is to handle collecting a high sampling rate of each sensor node. This challenge is similar to our system.

LogicQ [1] presents an underlying system for WSN to handle a logic-query that can be applied to make a WSN as a deductive database system.

Recently, the Microsoft's Data Center Genome project [9] presents a data center monitoring system using WSN that is similar to our work. However, they present only the collection part but do not cover the control part.

## 3. SENVM

SENVM is the first proposed monitoring and controlling environment system using WSN so far. The architecture of the system can be divided into three individual parts, i.e., a temperature sensor part, a temperature controller part, and a base station part. Generally, a base station collects temperature data from several sensors, determines condition in each area, and sends messages to controllers if temperature in any area needs to be changed, and so on. The architecture overview is shown in Figure 2. Wireless sensors are deployed by attaching to air-conditioners, servers, and a base station. Note that it is not necessary to attach wireless sensors to every single server, but the system will work perfectly if each rack has at least one sensor.

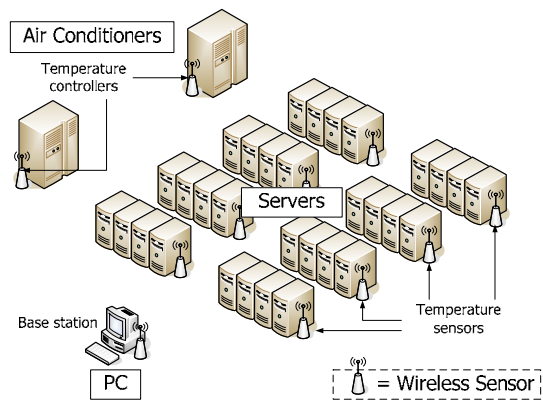

Figure 2. System architecture after deploying wireless sensors in a server room.

### 3.1 Temperature Sensor and Temperature Controller

A temperature sensor is used to measure environment at a specific area and send data back to the base station periodically. Ideally, the sensor should be placed in front of and behind every server, but practically, one temperature sensor for each server rack is enough.

We implement a temperature sensor with a Crossbow TelosB mote running on TinyOS 2.1, as shown in Figure 3a). The mote is powered by 2 AA-battery or USB port. This mote can measure temperature, humidity, and light intensity.

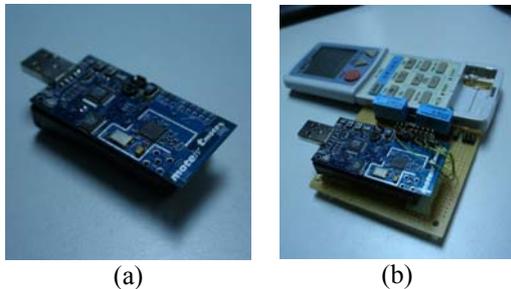

(a)　　　　　　　　(b)

Figure 3. a) Crossbow TelosB mote used as a temperature sensor and b) a universal air-conditioner remote controller equipped with a TelosB mote.

The temperature controller is also used Crossbow TelosB mote in implementation. Unlike the temperature sensor, the mote of temperature controller is wired with a universal air-conditioner remote controller shown in Figure 3b). When a server room is too hot or too cold, the base station sends a signal to the temperature controller and the signal is passed to an air-conditioner by the universal remote control.

Messages between motes are transferred in a tree-topology network. In real environment, the network condition may dynamically change over time. The wide range of network conditions affects the performance of a wireless sensor network. Therefore, to collect stream data from wireless sensor nodes efficiently, the data collection protocol must be reliable and robust. We implement Collection Tree Protocol (CTP) [5] that is a collection protocol providing more than 90% reliability and robustness.

Briefly, CTP forms a child-parent tree in a wireless sensor network. A sink node (base station node) is the root of a tree. Other source nodes in the network route their sensed data to the sink. At any specific time, each node has only one parent in a direction toward the root. However, CTP uses a link estimator to monitor a link quality; if a link quality is lower than expected threshold, a node can consider that link as a broken link and switch to another candidate link.

After tree formation is completed, each sensor node starts sampling temperature, humidity and light intensity and sends to the sink periodically. The sink, then, sends received data through a serial ported connected to a PC and triggers an applet running on the PC to store data to a database.

When temperature of air-conditioner must be changed, the sink receives a desired temperature command from the PC. Then, this command is routed through the tree toward a specific controller to change the controlled temperature

**3.2 Base Station**

The base station also plays an important role in the system. It sophisticatedly utilizes a fuzzy logic [16] to control temperatures for air-conditioners to be in an appropriate condition, i.e., 25ºC. Specifically, the fuzzy logic controls the system by iteratively retrieving temperature data and sending signals to controllers; in other words, the fuzzy logic can be considered as an iterative feedback system. For example, suppose desired temperature is set to be 25ºC. If a sensor currently measures as 26ºC, the system will determine and response to air-conditioner controller by 24ºC. If a temperature in a room is still the same, the system will send a signal with 23.5ºC (or lower), and so on. Figure 4 illustrates a concept of a fuzzy logic applied for SENVM.

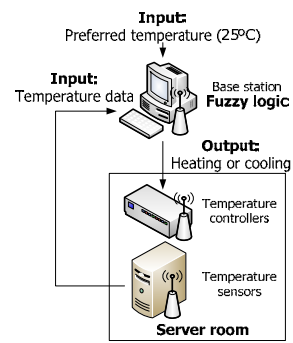

Figure 4. General idea of a fuzzy logic implemented in the base station.

The software architecture implemented at the base station is also designed to support high streaming rate, which requires well-organized index structure for fast retrieval and response. Specifically, the architecture consists of three layers, four components, i.e., wireless sensors, web browser, web server, and database, as shown in Figure 5.

At the base station, one wireless sensor operated by TinyOS is attached to be a gateway of the base station to connect to other wireless sensors, both temperature sensors and temperature controllers. Data are transferred from the wireless sensor through USB port, and the web server receives/sends via COM channel. When new incoming messages arrive, web server pre-processes, indexes, and stores data in the database. In every minute, after temperature sensors send temperature information to the base station, the web server computes the appropriate temperature for each air-conditioner, and sends messages to temperature control nodes.

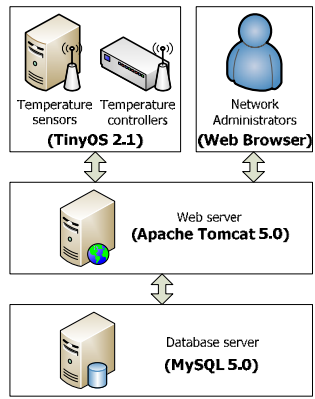

Figure 5. Three layers of software architecture at the base station.

Suppose that one node dies. The base station can detect the death node by checking that no message of that node arrived in the specific time (five minutes in our implementation). The dead node is then reported via the web server

The database used in the SENVM system can be any relational database (in our implementation, we use an open-source database system, MySQL 5.0). Generally, a primary key used to index data is a tuple of a node id and a timestamp; however, range query in time stamp makes processing not feasible since all retrieved data must be transferred to the web server. To make a fast retrieval, a timestamp is quantized in levels of minute, hour, and day. Figure 6 shows a diagram of a designed database table. Note that wireless sensors used in our implementation can collect additional information including air humidity and light intensity; we, therefore, design database for future used.

```
Air_Condition_Data
- id (int)
- timestamp (long)
- minute (long)
- hour (long)
- day (long)
- temperature (double)
- humidity (double)
- intensity (double)
```

Figure 6. Schema of a table used to store temperature data.

## 4. Demonstration

A prototype of SENVM is completely built and ready to deploy to a data center. Two deployment phases are planned, i.e., to test a monitoring function and then a controlling function. We success with our first phase deployment, and the second phase is going to deploy in this short time.

In the first phase, the prototype consists of four temperature sensors, and the base station. We use Crossbow TelosB motes [20] for all wireless sensor motes developed by UC Berkeley running on TinyOS 2.1, and the base station is powered by Intel Pentium 4 CPU 2.26GHz with 2 GB of RAM on Ubuntu 9.4. A room with two air-conditioners are set up and monitored by our SENVM.

The deployment site locates in two rooms resided in the $4^{th}$ building of Faculty of Engineering, Chulalongkorn University. The first room located on the $19^{th}$ floor is the small data center room of Department of Computer Engineering. Four temperature sensors are placed to monitor temperature, humidity, and light intensity. We call the server room as Room A for simple explanation. Figure 7 shows a room plan of the server room. In the room, there are two conditioners, four server racks, and two windows, where one large window is faced to the west. The base station is located in the second room on the $18^{th}$ floor, called Room B, where web server and database server are running.

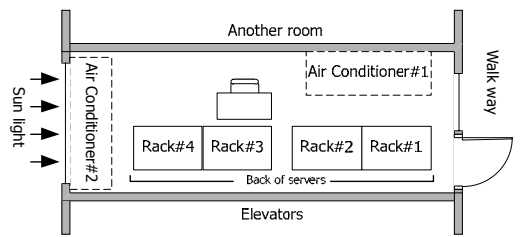

Figure 7. Rough blue print of the server room on the $19^{th}$ floor, where four server racks and two air conditioners are placed.

At Room A, we place four sensor nodes to monitor temperature, humidity and light intensity. Two sensor nodes (N10 and N12) are placed behind servers (hot zone), whereas, other two nodes (N9 and N11) are placed in front of servers (normal zone). An air conditioner is equipped on a ceiling in the top-left corner in the figure. At Room B, a sensor node (N0) acts as a root node and is connected with a PC. Another node (N1) acts as a relay node between two rooms. Figure 8 shows nodes at deployment site.

Arrow lines in Figure 9 indicate wireless links between each sensor pair. The shorter link is usually a better quality link. The arrow-end points to a parent node. Notice that some node has more than one parent. In fact, at any specific time, a node chooses only one parent from multiple candidate parents. When quality of a selected link becomes lower than a threshold, a node switches its parent to a candidate parent. For example, the node N11 can send data directly to the root node N0, however, when a link quality becomes low, it switches to send data to the node N1 instead. This provides more reliability and robustness than only one static parent.

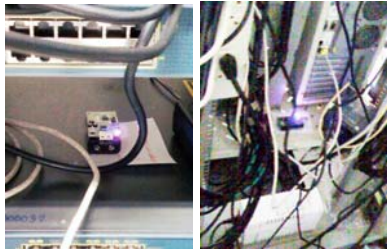

Figure 8. Temperature sensors *left)* in front of and *right)* behind servers at Room A.

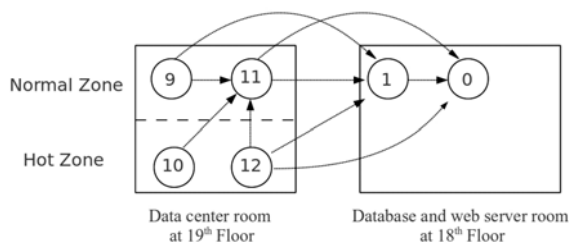

Figure 9. Connected links between sensor nodes.

Concretely, Figure 10 shows results of three-day monitoring of four temperature sensors, which are currently online [22][23]. As we can see, temperatures captured from several nodes are periodic. Therefore, we summarize maximum and minimum temperature for each node in Table 1. Additionally, many conclusions can be made from monitoring results. In day time, a load of each server increases, so a temperature at a hot zone (N10 and N12) significantly increases. Since a hot zone makes overall room temperature increase, air-conditioners try to decrease room's temperature as we can see from a temperature at day time of N9 and N11 in a normal zone decreases.

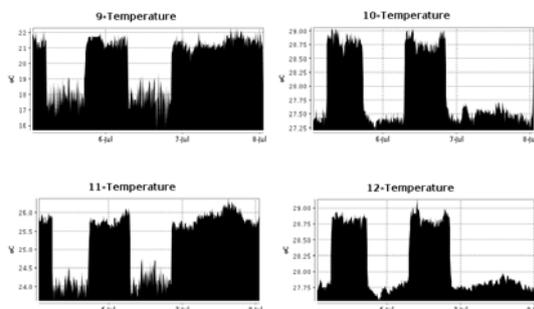

Figure 10. Temperatures of four temperature sensors during three-day monitoring.

Table 1. Statistics of temperature sensors.

| Node ID | Day time | Night time |
|---|---|---|
| | Avg ($^oC$) | Avg ($^oC$) |
| 9 | 19.72 | 21.87 |
| 10 | 28.79 | 27.37 |
| 11 | 24.03 | 25.84 |
| 12 | 28.81 | 27.46 |

Before we deploy our second phase, where a control part is implemented, we suggest moving server racks in order to receive cool air from air-conditioner equally. In other words, if a hot zone can receive more cool air, more efficient energy is utilized.

From our one year deployment, the SENVM system archives high performance in reliability and robustness. However, in early deployment period, we faced with some problems. First, when we deploy for the first time, we let all sensor nodes operated on the AA-battery power except the root node that operated on USB-line power. We notice that all non-root nodes drained power excessively and died in 3-4 days after deployment. The reason is our sampling rate is rather high (120 samples/minute). Consequently, each sensor node must turn on their radio transceiver most of time. The radio power consumption is the main factor of overall power consumption. Therefore, to make the system operates more than a year in the deployment environment; we changed all sensor nodes to operate on USB-cable power. This is viable in practice because most of servers in a data center room consist of multiple USB ports.

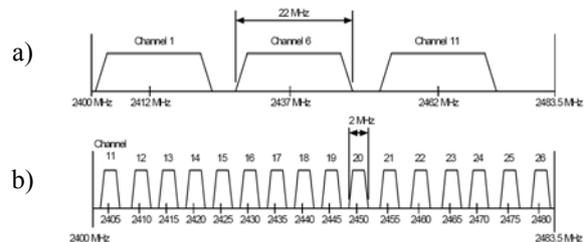

Figure 11. a) IEEE 802.11b North American channel selection (nonoverlapping) and b) IEEE 802.15.4 channel selection (2400 MHz PHY) [18]

Second, in deployment environment, we early notice many packet losses. Later, we found the reason is that the operating frequency is on the same with many IEEE 802.11 wireless access points. Figure 11 illustrates the operating frequencies of IEEE 802.11b/g that is used in wireless access points and IEEE 802.15.4 that is used in TelosB wireless sensor nodes. Both wireless access points and sensor nodes operate on the same 2.4GHz range with different gaps. IEEE 802.11b/g has three non-overlapping channels (channel 1, 6 and 11) and IEEE 802.15.4 has 16 non-overlapping channels (channel 11 - 26), however, both standards overlap each other in some channels. Therefore, we changed our wireless sensor nodes to operate on the last channel to avoid interference with wireless access points.

## 5. Conclusion and Future Work

Normally data centers in Thailand set air temperature around 20ºC, which leads to inefficient in energy utilization and also unnecessarily in electricity expenses since 25ºC is an appropriate operation temperature for a server. In this paper, we introduce a novel system, called SENVM, which is used to monitor and control air-conditioners in a data center using Wireless Sensor Networks (WSNs). With this SENVM, a solution to make a small data center to be "Green" is feasible. SENVM is designed to directly measure temperature at servers and send a control signal to air-conditioners whether a server is too hot or too cold; therefore, SENVM can make sure that temperature at servers will be in an appropriate condition all time.

In addition, this system can be extended to a larger data center and implemented with various time series mining techniques such as anomaly detection [13] and pattern matching [11][12].